# A LARGE LANGUAGE MODEL APPROACH TO CLASSIFY FLAKINESS IN C++ PROJECTS

Xin Sun, Daniel Ståhl, and Kristian Sandahl

Department of Computer and Information Science, Linköping University, Linköping, Sweden

***ABSTRACT***

*The role of regression testing in software testing is crucial as it ensures that any new modifications do not disrupt the existing functionality and behaviour of the software system. However, the presence of Flaky Tests undermines the reliability of regression testing results. In this paper, we propose an LLM-based approach for classifying the root cause of identified flaky tests in C++ projects at the code level. We compile a comprehensive collection of C++ project flaky tests sourced from GitHub. We finetune Mistral-7b, Llama2-7b and CodeLlama-7b models on the C++ dataset and an existing Java dataset and evaluate the performance. The results indicate that our models exhibit varying performance on the C++ dataset, while their performance is comparable to that of the Java dataset. Our results demonstrate the exceptional capability of LLMs to accurately classify flakiness in C++ and Java projects, providing a promising approach to enhance the efficiency of debugging flaky tests in practice.*

## 1. INTRODUCTION

Regression testing is deployed to prevent code changes from causing any disruptions to the existing functionality of the software[1]. The presence of flakiness in tests can have an impact on the results of regression testing and undermine its reliability [2].

Flaky tests have long been identified as a challenge in software testing [3, 4]. A flaky test is a test case that passes or fails randomly without any change made to the test code or the Item Under Test (IUT) [5], resulting in multiple problems in the software development process. Flaky tests may not consistently reveal faults in the IUT, leading to uncertainty about whether the problem originates from the test case or the IUT itself. Failures caused by flaky tests are often attributed to issues with test execution or the inherent design of the test case [6]. Flaky tests can affect automatic builds with false signals and cause undesirable delays in the Continuous Delivery (CD) [7, 8]. An obvious consequence is that flaky tests can cause developers to waste time debugging, which decreases their trust in the test suite [4] and negatively impacts their productivity [9, 10].

A study of Google's tests reported that 41% out of 115,160 test targets show some kind of flakiness [11]. Also, a previous work [12] reported that about 4.6% of the tests in five Microsoft projects are flaky. One common approach to finding a flaky test is to run the test cases many times, which is inefficient and time-consuming [3, 13, 14]. It is also hard to determine the appropriate number of reruns to find discrepancies in outputs [15].

Researchers have proposed many methods to detect flaky tests with fewer resources. IDFlakies [16] suggested running the test cases in a different order, which can detect the flakiness with fewer reruns. Zhang et al. [17] studied several real-world dependent tests, then proposed and compared four algorithms to detect dependent tests in a test suite. Pinto et al. [18] detected the flaky tests in regression test suites using various machine learning algorithms. King et al. [19] leveraged a Bayesian network to classify and predict flaky tests. Besides detecting flaky tests, some research focused on classifying the category of flaky tests. Luo et al. [20] studied several open-source projects and classified the rootcauses of flaky tests into ten categories. Lam et al. [16] published a study where they classified flaky tests to be order-dependent or non-order dependent.

Recently, Large Language Models (LLMs) have become popular because of their excellent performance in text understanding and generation. Significant research has focused on utilizing LLMs to address various challenges, especially code-related tasks. However, the application of LLMs for classifying test flakiness remains a relatively novel area of exploration. For example, FlakyCat [7], a CodeBERT-based multi-class classifier with Few-Shot Learning, categorizes flaky tests based on their root causes of them. Furthermore, most research focused on the flaky tests in Java and Python projects, not C++ projects. C++ is one of the most used programming languages, and it is a good choice for embedded, resource-constrained programs [21]. Hence, it is necessary to investigate the method to classify the flakiness in C++ flaky tests.

To address this, we leverage three different LLMs and fine-tune them on a C++ flaky test dataset to investigate the efficacy of flaky test classification beyond the scope of previously studied languages while concurrently comparing the performance of multiple state-of-the-art large language models. We will address the following research questions in this paper:

**RQ1: How accurately can our approach classify the flakiness categories of C++ projects compared to the existing Java dataset?**

In this research question, we aim to study the performance of different models in classifying the flakiness in Java and C++ projects. Then, we analyze the F1 score, accuracy, precision, and recall to evaluate their abilities.

**RQ2: How does our approach compare to previous work for the classification of flakiness in Java flaky tests?**

This research question examines the performance of different LLMs in classifying flakiness in Java projects and then compares the results to previous work. In our study, we use the Java dataset of FlakyCat [7], so we compare our results with FlakyCat.

The paper makes the following contributions:

– We present a C++ flaky test dataset for our project and future research. To the best of our knowledge, this represents the first publicly available dataset of C++ flaky tests. The dataset is valuable for further research on C++ flaky tests.
– We propose a method to classify the flakiness of C++ and Java flaky tests.
– We compare the capability of our selected models to classify flakiness categories in Java and C++ flaky tests, offering suggestions for model selection in future research.

The datasets used in this study and the scripts are publicly available in our GitHub repository (https://github.com/PELAB-LiU/FlakyClassifier) to facilitate reproducibility.

The rest of this paper is organized as follows: Section 2 presents the background and related works. Section 3 presents our approach. Section 4 shows the results of our approach. Section 5 discusses our results. Section 6 shows the threats to validity. Finally, Section 7 concludes the paper and suggests future work.

## 2. BACKGROUND

This section presents the background of our study. We list the root causes of flaky tests and related works in this area.

### 2.1. Root Causes of Flaky Test

It is vital to understand the root causes of flaky tests before fixing them. Researchers and practitioners conducted many studies to determine the root cause of test flakiness [10,20]. Luo et al. [20] conducted an extensive study of flaky tests by analysing the commit history of all projects from the Apache Software Foundation. They filtered and analysed the commits, resulting in 201 commits suitable for inspection. After analysis, the root causes were divided into ten categories: Async Wait, Concurrency, Test Order Dependency, Resource Leak, Network, Time, IO, Randomness, Floating Point Operations, Unordered Collections. Their results showed that most of the flaky tests in their dataset were caused by Async Wait, Concurrency and Test Order Dependency. Then, Eck et al. [10] reported four uncovered causes of test flakiness in their study. The study suggested that Too Restrictive Range, Test Case Timeout, Platform Dependency and Test Suite Timeout could also be the root cause of some test flakiness. In our study, we constructed our dataset following the categories they set and fine-tuned the LLMs to classify the flaky tests into the categories mentioned above. Table 1lists the root causes and their brief definitions used in our study.

### 2.2. Related Work

A large language model is a statistical model with billions of parameters, and it is trained to predict the next few words in a sequence [22]. These models are pre-trained on vast amountsof text data and have demonstrated powerful performance in a wide range of Natural Language Processing (NLP) tasks, including language translation, text generation, text classification and code generation [23–26]. Inspired by the excellent performance on code understanding of LLMs, practitioners also tried to deploy them to flakiness classification.

Aklli et al. [7] proposed FlakyCat, a CodeBERT-based approach to classify flaky tests based on their root cause category. They also leveraged Siamese networks [27], which consist of a pair of networks that share weights and are designed to compute similarities between elements to classify the flakiness. Fatima et al. [28] utilized LLMs to classify flaky tests based on how they were fixed. They constructed a dataset of fix categories based on existing datasets. Then, they built prediction models based on CodeBERT and UniXcoder and trained them on the dataset they had created. The approach can output repair advice for flaky tests. Their work alleviates the issues of flaky tests to some extent. However, the datasets used in previous work only contain Java and Python projects, which is insufficient for the software testing field.

Table 1. Root Causes of Flaky Tests

| Root Causes | Definitions |
|---|---|
| **Async Wait** | A test makes an asynchronous call but does not properly wait for the result to become available before using it. |
| **Concurrency** | The test non-determinism is due to different threads interacting in a non-desirable manner. |
| **Test Order Dependency** | The test outcome depends on the order in which the tests are run. |
| **Resource Leak** | The test does not properly manage one or more of its resources, such as memory, resulting in test failures. |
| **Network** | The test depends on network connections, and it fails when the network is unstable. |
| **Time** | The test relies on the system time, which introduces non-deterministic failures. |
| **I/O** | The test has I/O operations which may cause flakiness. |
| **Randomness** | Random numbers can make some tests flaky if all possible values are not properly addressed in the test. |
| **Floating Point Operations** | The test with floating point operations can be non-deterministic. |
| **Unordered Collections** | The test outcome can become non-deterministic if you assume the elements are returned in a particular order. |
| **Too Restrictive Range** | The test has a restrictive or insufficient assertion, which may cause non-deterministic behaviour. |
| **Test Case Timeout** | The test does not produce an output for a fixed amount of time and leads to a flaky outcome. |
| **Platform Dependency** | The test fails consistently on a specific platform but passes on other platforms. |
| **Test Suite Timeout** | The test suite non-deterministically times out, not a single test case. |

## 3. RESEARCH METHOD

In this section, we introduce an approach based on LLMs to predict the flakiness in flaky test cases. We present our datasets and the fine-tuning of the LLMs. Additionally, we illustrate the evaluation of our approach.

### 3.1. Datasets Construction

Data is a vital source of LLM training. A high-quality dataset makes the training more accessible and improves the performance of the models. In our approach, we fine-tuned the LLMs on two datasets: one is C++, and the other is Java. We collected C++ flaky tests from open-source projects on GitHub and applied data augmentation to them to get the C++ dataset. The Java dataset we used is from FlakyCat [7].

*C++ Dataset*

In our approach, we had to collect a set of C++ flaky tests to construct a dataset for fine-tuning LLMs. We decided to extract the flaky tests from open-source projects on GitHub.

Initially, we searched the keyword “flaky” on GitHub, restricting the programming language to “C++”. This search yielded over 58,000 results. However, the results contained many irrelevant items. To further refine the results, we applied the “Issues” filter on the website, limiting the search to issues with “flaky” in their titles. Among these results, some were newly identified and remained unresolved. It is hard to assign a category for those issues without any comments from the developers. Thus, we removed these issues by filtering out those not labelled as “Closed”.

The previous steps yielded a set containing issues about C++ flaky tests. Next, we filtered out issues that were hard to classify or duplicated. Additionally, some flaky tests were fixed by modifying the production code or other associated files, so we excluded these tests from our data to focus solely on the ones caused by external factors or test-specific issues. Finally, we got a dataset containing 55 C++ flaky tests and comments from the developers. Based on the comments from the developers and the test code, we manually categorized each flaky test according to Table 1.

### *Data Augmentation*

The collected dataset was too small for LLM fine-tuning. Thus, we had to deploy data augmentation to obtain more flaky tests. Inspired by [7, 29–31], we leveraged GPT-4 and Synthetic Minority Over-sampling Technique (SMOTE) [32] to augmentour C++ dataset. SMOTE is an over-sampling technique that generates extra synthetic samples from the minority class. Similar to the approach used in FlakyCat, in our study, we only changed the variable names and constants for each test case from the C++ dataset and added declarations of unused variables. In this way, we ensured that the underlying flakiness of the tests remained unaffected, allowing the models to learn relevant information about test flakiness without being influenced by superficial code changes. Unlike FlakyCat [7], we prompted GPT-4 to achieve this task. GPT-4 performs well in code understanding and can easily replace the variable names and constants with similar words instead of introducing meaningless characters. A meaningful variable name could help the LLMs used in our study to understand the tests better and learn more meaningful patterns related to test flakiness. We set the prompt as follows:

Here are some C++ flaky test cases. Please use the SMOTE method to augment the given code. Please mutate only the variable names, constants or test method names or add declarations of unused variables. Do not influence the flakiness of the code. For each given code, please return five augmented examples to me.

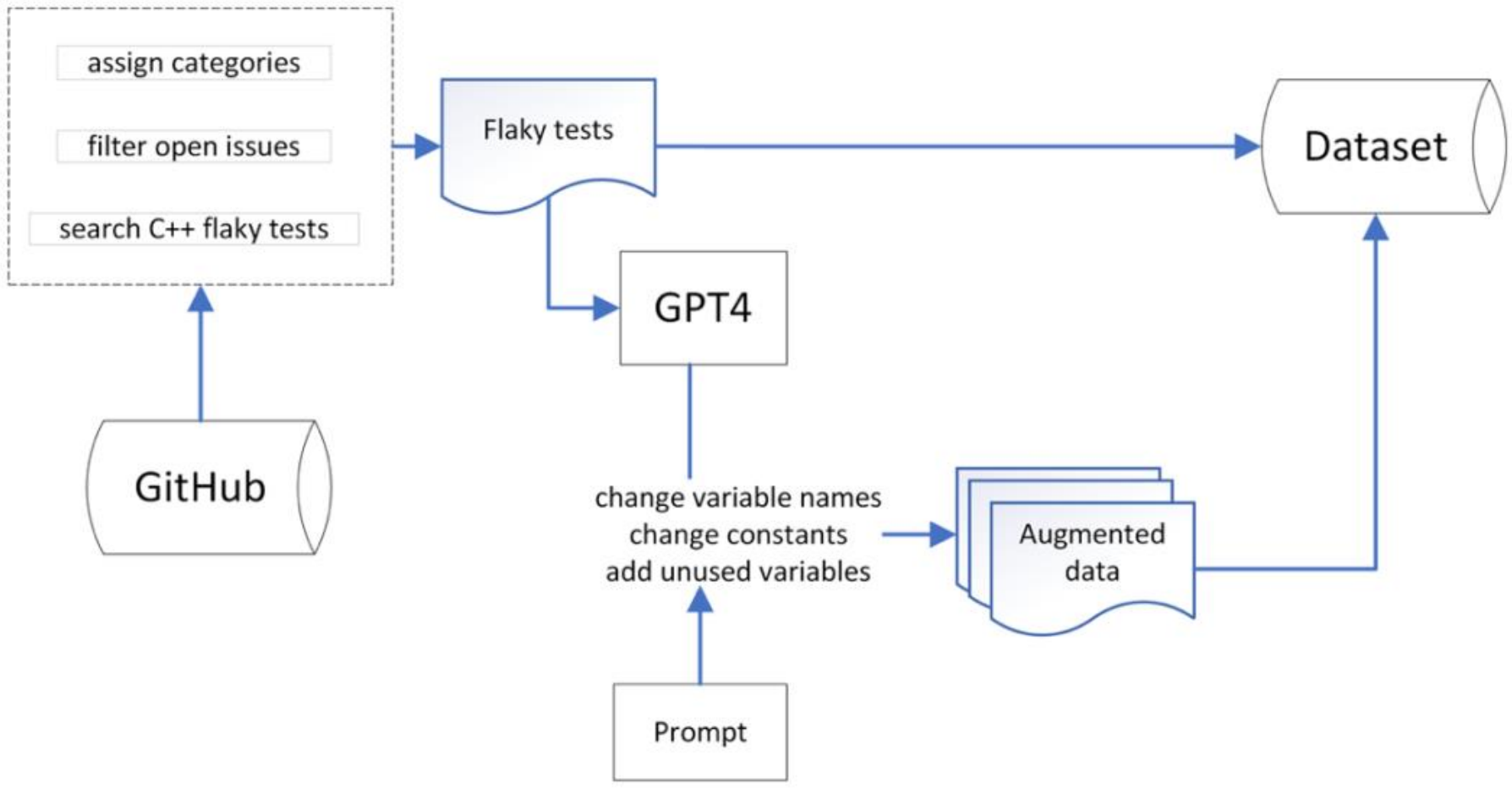

Figure 1. The overall process of constructing our C++ dataset

Considering the max length limit of GPT-4, we input one test each time. After the augmentation, we filtered some of the generated tests if the number of text variants was fewer than three. This ensured that the augmented dataset maintained sufficient diversity in the test cases. Finally, we combined the augmented dataset with the collected dataset and got a dataset containing 362 flaky test cases.

Figure 1shows how we built our C++ dataset.

Figure 2shows an example of our collected flaky test, and the corresponding augmented flaky test. In the left figure, the counter **executing [pool id]** is modified without proper synchronization, which can lead to race conditions if multiple threads update it concurrently. After data augmentation, the names of different variables were changed. For example, **auto job func** was replaced by **auto task function**, GPT4 only changed the name but didn't change **auto**, which is a keyword used for type inference in C++. Furthermore, GPT-4 added an unused variable at the beginning of the code, which is **unusedVar1**. The flakiness was not affected. With this, we also avoided using meaningless variable names, which may help the LLMs understand the tests better during classification.

```
1
2      auto job_func = [&] (
     AsyncLoader & loader, const
     LoadJobPtr & self)
3      {
4          auto pool_id = self->
     executionPool();
5          executing[pool_id]++;
6
7          if (executing[pool_id] >
     max_threads[0])
8              boosted = true;
9          ASSERT_LE(executing[
     pool_id], max_threads[pool_id
     ]);
10
11         if (--left == 0)
12         {
13             std::unique_lock lock
     {prioritization_mutex};
14             for (const auto & job
      : jobs_to_prioritize)
15                 loader.prioritize
     (job, 1);
16
17         }
18         std::shared_lock lock{
     prioritization_mutex};
19         t.randomSleepUs(100, 200,
      100);
20
21         ASSERT_LE(executing[
     pool_id], max_threads[pool_id
     ]);
22         executing[pool_id]--;
23
24     };
```

```
1      int unusedVar1 = 42;
2          auto task_function = [&]
     (AsyncLoader & loader, const
     LoadJobPtr & self_task)
3          {
4              auto pool = self_task
     ->executionPool();
5              running_tasks[pool
     ]++;
6              if (running_tasks[
     pool] > thread_limits[0])
7
     concurrencyBoosted = true;
8              ASSERT_LE(
     running_tasks[pool],
     thread_limits[pool]);
9
10             if (--jobs_left == 0)
11             {
12                 std::unique_lock
     lock{mutex_for_prioritization
     };
13                 for (const auto &
      job : jobs_to_prioritize)
14                     loader.
     prioritize(job, 1);
15             }
16             std::shared_lock lock
     {mutex_for_prioritization};
17             testInstance.
     randomSleepUs(100, 200, 100);
18
19             ASSERT_LE(
     running_tasks[pool],
     thread_limits[pool]);
20             running_tasks[pool
     ]--;};
```

Figure 2. Code snippet before augmentation (left) and Code snippet after augmentation (right)

Finally, we got a dataset containing 362 C++ flaky tests. Table 2 shows the distribution of our C++ dataset. **Original** means the data were extracted from flaky tests of GitHub projects. **Augmented** shows the data after data augmentation. **Final dataset** is the combination of them.

***Java Dataset***

The approach we adopted involved the utilization of two distinct datasets of different programming languages. One dataset was constructed for C++, while the other was initially used by FlakyCat [7]. The Java dataset comprises flaky tests from previous work [33–35], GitHub projects and data augmentation. We performed a filtering process to eliminate samples with missing labels, resulting in the refined Java dataset. The Java dataset contains 1287 Java flaky tests. Some categories have less than 15 examples, and it is not feasible for models to learn from very few examples. Thus, we removed those categories, which are **Float point operation, I/O, Platform dependency**, and **Too restrictive range**, from the dataset before fine-tuning and evaluation. Finally, we got a dataset containing 1287 Java flaky tests.

Table 3 illustrates the information of our two datasets. Both datasets are not distributed evenly across categories of flakiness; this means we have imbalanced datasets. As illustrated above, we filtered categories with less than 15 flaky tests in the Java dataset. Thus, in our experiment, we did not use those categories of each dataset to train and evaluate the models.

Table 2. The distribution of our C++ dataset

| **Category** | **C++ Dataset** | | |
|---|---|---|---|
| | **Original** | **Augmented** | **Final dataset** |
| Async wait | 12 | 78 | 90 |
| Concurrency | 7 | 50 | 57 |
| Time | 18 | 77 | 95 |
| Unordered collections | 2 | 20 | 22 |
| Float point operation | 5 | 32 | 37 |
| I/O | 2 | 18 | 20 |
| Randomness | 6 | 18 | 24 |
| Too restrictive range | 3 | 14 | 17 |
| **Total** | 55 | 307 | 362 |

Table 3. The information of our datasets

| **Category** | **C++ Dataset** | **Java Dataset** |
|---|---|---|
| Async wait | 90 | 376 |
| Concurrency | 57 | 144 |
| Time | 95 | 122 |
| Unordered collections | 22 | 153 |
| Randomness | 24 | 50 |
| Network | - | 93 |
| Test Order Dependency | - | 291 |
| Resource Leak | - | 28 |
| Platform dependency | - | 6 |
| Float point operation | 37 | 9 |
| I/O | 20 | 6 |
| Too restrictive range | 17 | 9 |
| **Total** | 362 | 1287 |

## 3.2. Fine-Tuning

In this section, we describe how we fine-tuned different LLMs to predict and classify flakiness. Currently, the two prominent approaches to enhance the capability of LLMs are Finetuning and Retrieval Augmented Generation (RAG). RAG uses external data to augment the prompt, while fine-tuning injects the additional knowledge into the base model [36]. Both of them are highly efficient in adapting LLMs to specific domains. In our study, considering the flakiness classification is less popular among code generation, completion and code review, we decided to use fine-tuning instead of RAG. The LLMs need to learn new skills specific to this domain. Fine-tuning can make LLMs provide more precise and succinct responses without needing to perform expensive retrieval steps [36], and make the LLMs more efficient for repeated use.

Our approach involves fine-tuning three language models: Llama2-7b, CodeLlama-7b and Mistral-7b. Llama2 [37] is a family of pre-trained and fine-tuned large language models. They have several versions depending on the number of parameters, ranging from 7 billion to 70 billion. Llama2 models outperform many open-source models on most benchmarks and can be a suitable substitute for closed-source models like GPT-4 [31]. CodeLlama, a family of code-specialized Llama2 models, shows state-of-the-art performance in programming tasks, especially code generation tasks. Mistral-7b [38] is also a large language model. It leverages grouped-query attention (GQA) and sliding window attention (SWA). Mistral7b outperforms the Llama2-13b model across all tested benchmarks and approaches the performance of CodeLlama-7b in coding tasks [38].

To fine-tune the models, we split each of our datasets into the training subsets and the evaluation subsets. The three models were fine-tuned on the training datasets and evaluated on the evaluation datasets, respectively. We allocated 75% for training and 25% for evaluation, which is the same with FlakyCat [7].

The experiments were run on a Nvidia DGX A100 compute node of Berzelius, the premier AI/ML cluster at NSC. For the fine-tuning strategy, we used Low-Rank Adaptation (LoRA). LoRA [39] is an approach that can freeze the pre-trained model weights and inject trainable rank decomposition matrices into each layer of the transformer architecture. In practice, you can use lora rank, lora alpha and lora dropout to control the performance of the method. Lora rank specifies the rank of the low-rank matrices. Lora alpha is the scaling factor to the LoRA updates. Lora dropout helps prevent overfitting during finetuning. Using LoRA can significantly reduce the number of trainable parameters, which can accelerate the training process and reduce the computation resource requirements.

In the fine-tuning process, we set the same value across all experiments for some finetuning parameters based on the work from Li et al. [40]. They fine-tuned Llama2-7b to do similar tasks. Following their settings, we configured the LoRA settings lora rank,lora alpha and lora dropout to 12, 32 and 0.1, respectively.

We set the epoch to 10 so the model can go through the dataset 10 times to improve its learning. Both the training and evaluation batch size were eight, which means the model processes eight samples at a time.“Max length” refers to the maximum number of tokens the model can process in a sequence. Since most of our test cases are short, we set it to 512. The learning rate for training is set to 5e-5 with a decay rate of 0.01, which accelerates convergence and helps avoid local minima.

## 3.3. Evaluation

To evaluate the performance of our classifier, we use several standard evaluation metrics, including precision, recall, F1 score, and accuracy. Since our datasets have imbalanced categories, the F1 score is the primary metric for evaluation. In the evaluation, we used weighted averaging [41], which is better for imbalanced datasets. Weighted averaging is a method used to aggregate precision, recall and other performance metrics in multi-class classification problems. It considers the size of each category when calculating the overall metric. By doing this, it ensures that larger categories have a greater influence on the final score. In addition to the performance metrics, we evaluated the model's performance after every epoch. This evaluation strategy provides a better understanding of the fine-tuning process, and we can monitor the improvement of the models after every epoch.

# 4. RESULTS

In this section, we present the results of our experiments and answer our research questions according to the results. We show the accuracy, the weighted average of the F1 score, precision and recall of every model after every training epoch in Table 4. We also presentthe curve of accuracy and F1 score of our models on the two datasets in Fig 2 and Fig 3. Finally, Table 5 presents the precisions and recalls of each category in our dataset. Fig 4 and Fig 5 show the comparisons of our models' precision and recall for the four shared categories on our datasets.

**RQ1: How accurately can our approach classify the flakiness categories of C++ projects, comparing to the existing Java dataset?**

To evaluate our approach, we fine-tuned three models on both our C++ dataset and the Java dataset of FlakyCat [7]. We used the highlighted categories from each dataset to train and evaluate the three models, with the overall results presented in Table 4.

Table 4 provides a detailed summary of the performance of the models. The results indicate that the Mistral-7b model achieved a perfect classification on the C++ dataset, obtaining a score of 1.0 across all metrics. The Llama2-7b model, by comparison, reached a score of 0.90 for all the metrics. However, the CodeLlama-7b model only achieved an F1 score of 0.79 and an accuracy of 0.82 on the C++ dataset, which is the lowest among all the results on both datasets.

Table 4. Results of each model on our datasets

| | Model | F1 | Accuracy | Precision | Recall |
|---|---|---|---|---|---|
| **Java** | **Mistral-7b** | 0.85 | 0.87 | 0.86 | 0.84 |
| | **Llama2-7b** | 0.89 | 0.89 | 0.89 | 0.89 |
| | **CodeLlama-7b** | 0.86 | 0.84 | 0.87 | 0.87 |
| **C++** | **Mistral-7b** | 1.0 | 1.0 | 1.0 | 1.0 |
| | **Llama2-7b** | 0.90 | 0.90 | 0.90 | 0.90 |
| | **CodeLlama-7b** | 0.79 | 0.82 | 0.78 | 0.79 |

On the Java dataset, all three models showed comparable performance, with the Llama2-7b model slightly outperforming the other two models. It achieved an F1 score of 0.89. Mistral-7b achieved a slightly lower F1 score than CodeLlama-7b, but it obtained higher accuracy on the Java dataset. The results suggest that the Llama2-7b model provides a balanced performance across datasets.

Regarding weighted precision and recall, the Llama2-7b model demonstrated satisfactory performance on both datasets. It obtained scores of 0.89 for precision and recall on the Java dataset, which increased to 0.90 on the C++ dataset. The CodeLlama-7b and the Mistral-7b model achieved comparable precision and recall on the Java dataset, both around 0.87. However, on the C++ dataset, the results of the CodeLlama-7b model dropped below 0.80.

Figure 3shows the three models achieve similar performance on the Java dataset. All the curves start above 0.40 and stabilize at around 0.80 after six epochs. As shown in Figure 4, our models displayed different performance when classifying the flakiness in the C++ dataset. The Mistral-7b model outperformed the other two models throughout all epochs. The other two models exhibited similar performance until epoch four, after which the Llama2-7b model began to outperform the CodeLlama-7b model.

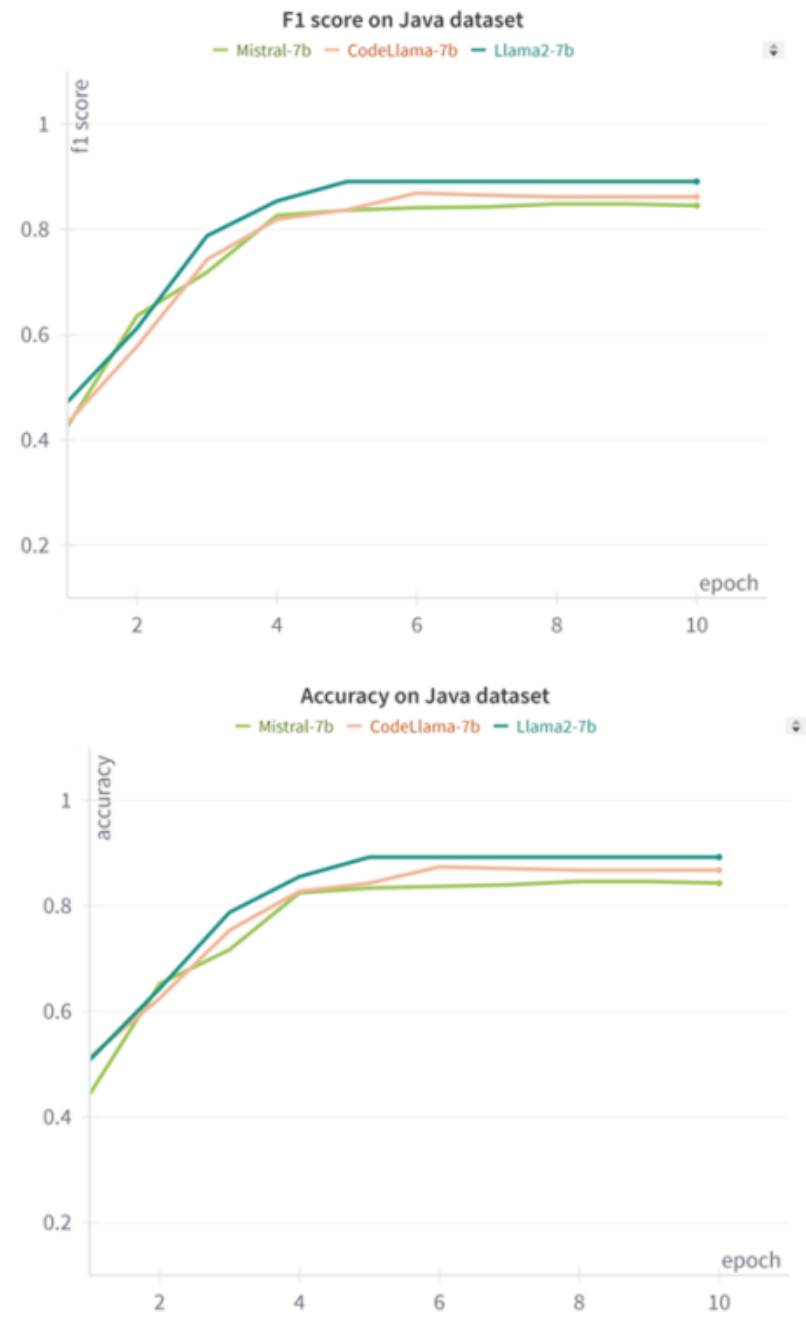

Figure 3. Results on Java dataset: We evaluated the models after every epoch. The curves show the performance of each model after every epoch

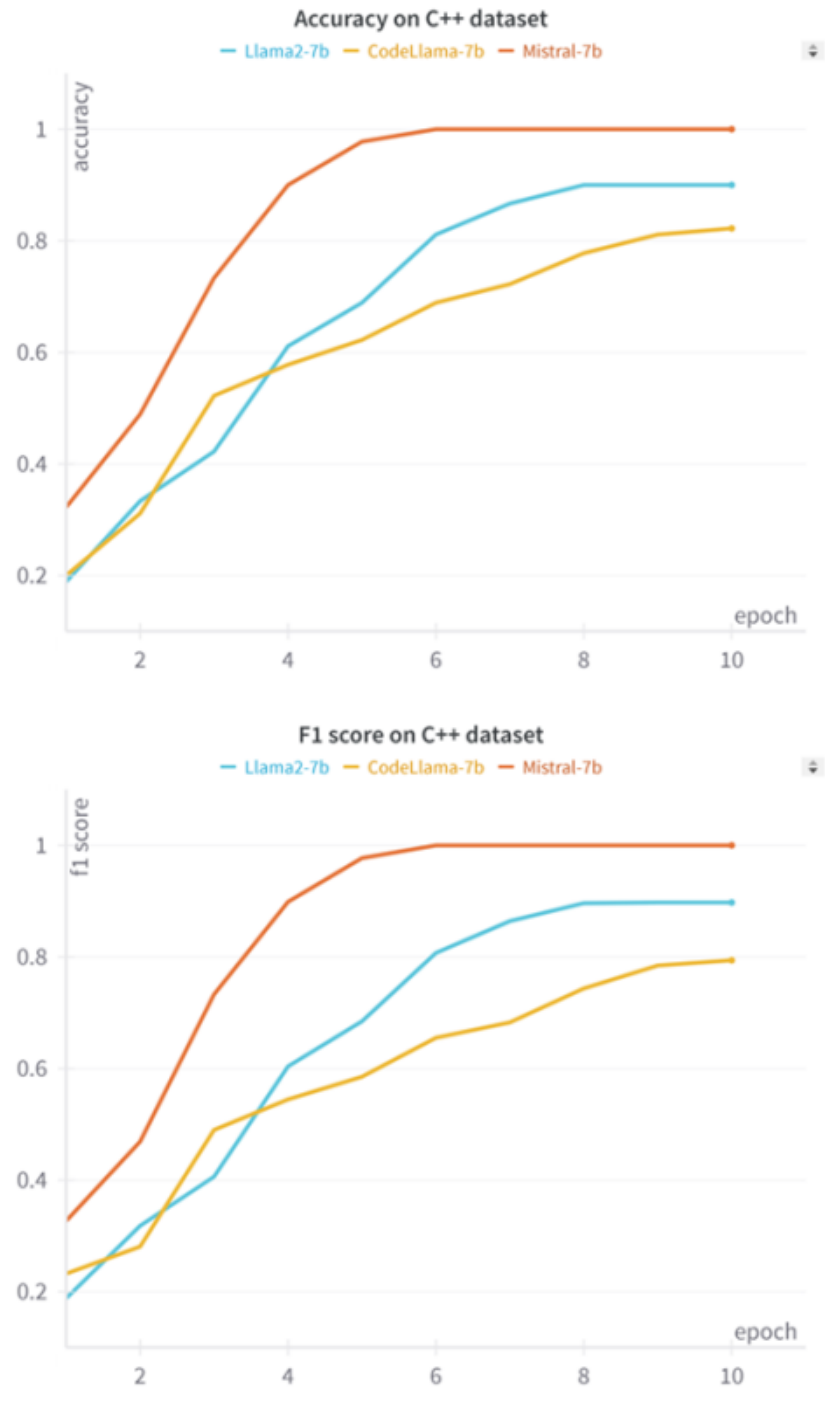

Figure 4. Results on C++ dataset: We evaluated the models after every epoch. The curves show the performance of each model after every epoch.

To understand the details of the precision and recall, we also examined the performance for each category in our datasets. We compared the results of the common categories between both datasets, which are highlighted in Table 5. Among all the categories, the Mistral-7b model achieved a precision and recall of 1.0 for the C++ dataset. For the CodeLlama-7b model, we observed significant variations between precision and recall insome categories. For instance, the precision of Randomness was 1.0, while the recall was 0.25. This also made it achieve a lower F1 score than the other two models. Figure 5 and Figure 6 show a clear comparison among the models. The Llama2-7b model performed well in recognizing **Async wait** and **Time** but underperformed in classifying **Concurrency** and **Randomness** in the Java dataset. Additionally, the precisions of the CodeLlama-7b model were comparable to those of the other models; however, its recalls varied significantly across different categories. This variation in recall was the primary factor contributing to its lower F1 score.

Table 5. Precision and recall of each model. The P represents for Precision, and the R represents for Recall.

| **Category** | **Java** | | | | | | **C++** | | | | | |
|---|---|---|---|---|---|---|---|---|---|---|---|---|
| | Mistral | | Llama2 | | CodeLlama | | Mistral | | Llama2 | | CodeLlama | |
| | P | R | P | R | P | R | P | R | P | R | P | R |
| Async wait | 0.87 | 0.88 | 0.89 | 0.94 | 0.81 | 0.97 | 1.0 | 1.0 | 0.96 | 0.96 | 0.77 | 0.97 |
| Concurrency | 0.81 | 0.66 | 0.88 | 0.74 | 0.84 | 0.71 | 1.0 | 1.0 | 0.88 | 1.0 | 0.79 | 0.79 |
| Time | 0.85 | 1.0 | 1.0 | 1.0 | 0.90 | 0.97 | 1.0 | 1.0 | 0.89 | 0.94 | 0.80 | 0.94 |
| Test case timeout | 0.50 | 0.80 | 0.67 | 0.80 | 0.88 | 0.70 | - | - | - | - | - | - |
| Unordered collections | 0.95 | 0.93 | 0.90 | 0.96 | 0.91 | 0.96 | - | - | - | - | - | - |
| Float point operation | - | - | - | - | - | - | 1.0 | 1.0 | 0.87 | 0.93 | 0.87 | 0.93 |
| Hash operation | - | - | - | - | - | - | 1.0 | 1.0 | 0.80 | 0.67 | 1.0 | 0.17 |
| I/O | - | - | - | - | - | - | 1.0 | 1.0 | 1.0 | 0.33 | 0 | 0 |
| Randomness | 0.82 | 0.75 | 0.75 | 0.7 | 0.73 | 0.67 | 1.0 | 1.0 | 1.0 | 0.75 | 1.0 | 0.25 |
| Too restrictive range | - | - | - | - | - | - | 1.0 | 1.0 | 1.0 | 1.0 | 0.50 | 0.50 |
| Network | 0.92 | 0.60 | 0.88 | 0.75 | 1.0 | 0.55 | - | - | - | - | - | - |
| Test Order Dependency | 0.84 | 0.91 | 0.91 | 0.93 | 0.95 | 0.91 | - | - | - | - | - | - |
| Resource Leak | 1.0 | 0.60 | 1.0 | 0.70 | 1.0 | 0.50 | - | - | - | - | - | - |

**RQ2: How does our approach compare to previous work for the classification of flakiness in Java flaky tests?**

In our approach, we fine-tuned three LLMs to perform an eight-class classification task on the Java dataset. For our datasets, the Java dataset comes from FlakyCat [7], which deployed the CodeBERT model and Few Shot Learning to classify the flakiness in Java. FlakyCat compared the performance of the CodeBERT-based model with traditional machine learning classifiers. Since we used the Java dataset of FlakyCat, we took the results of the CodeBERT-based model as a baseline for comparison.

Our findings indicate that the three LLMs we fine-tuned outperform the CodeBERT-based method of FlakyCat. As shown in

Table 6, in our results, the Llama2-7b model achieved the highest F1 score of 0.89, surpassing FlakyCat's result by 0.16. Moreover, the Llama2-7b model attained precision and recall scores of 0.89 on the Java dataset. Meanwhile, the CodeBERT-based method of FlakyCat got 0.74 and 0.73 for precision and recall, respectively.

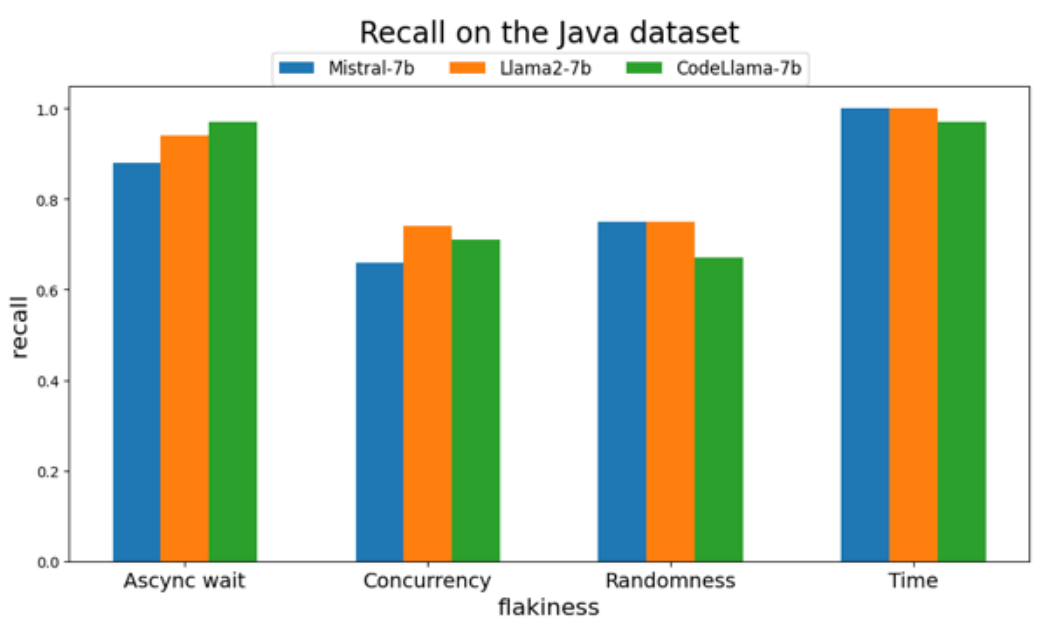

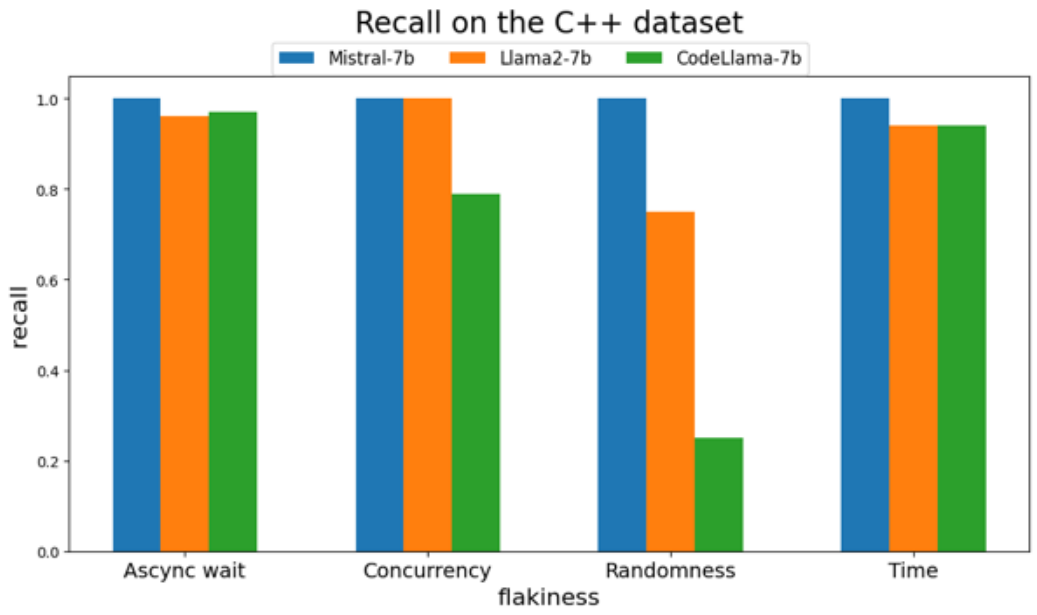

Figure 5. Comparison of model recall scores for shared categories on the Java and C++ datasets

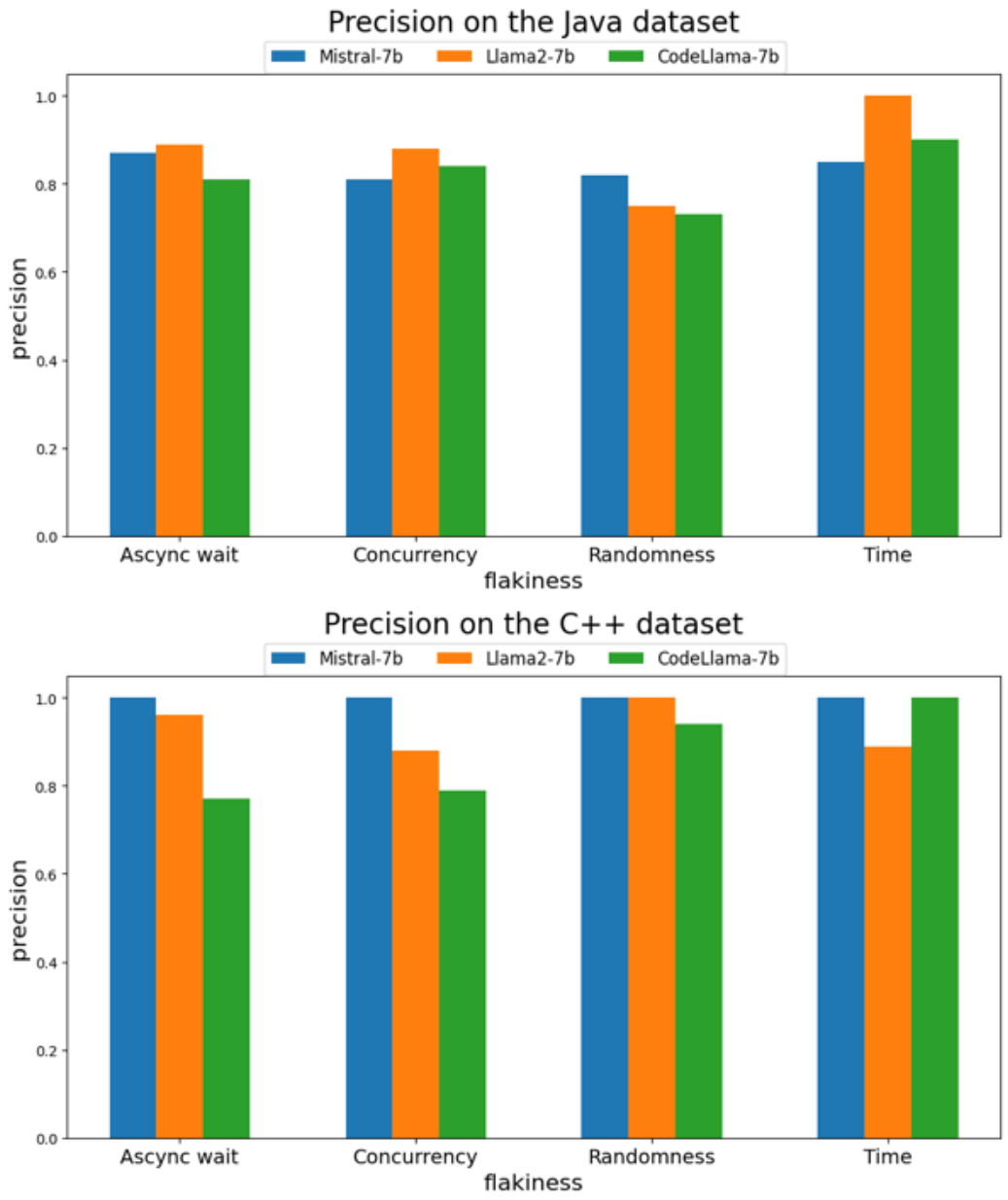

Figure 6. Comparison of model precision scores for shared categories on the Java and C++ datasets.

Table 6. Comparison of results between our study and FlakyCat.

| | **Model** | **F1 score** | **Precision** | **Recall** |
|---|---|---|---|---|
| **Our Study** | Mistral-7b | 1.0 | 1.0 | 1.0 |
| | Llama2-7b | 0.90 | 0.90 | 0.90 |
| | CodeLlama-7b | 0.79 | 0.78 | 0.79 |
| **FlakyCat** | CodeBERT-based | 0.73 | 0.73 | 0.73 |

## 5. DISCUSSION

Our results indicated that various models exhibited different capabilities in classifying flakiness across different programming languages. The Mistral-7b model demonstrated excellent performance on the C++ dataset, achieving a score of 1.0 across all our metrics.However, on the Java dataset, it only scored 0.85 for the F1 score, 0.87 for accuracy and 0.84 for recall. Similarly, the CodeLlama model exhibited a comparable scenario. It achieved an F1 score of 0.86 on the Java dataset, but the F1 score dropped to 0.79 on the C++ dataset. One possible explanation is that the Mistral model's training data may have included a larger proportion of C++ code, which enhanced its performance on C++ datasets. In contrast, the CodeLlama model was trained in multiple programming languages to offer intelligent suggestions and completions. This could explain why the Mistral-7b model performed better in understanding C++ codes. Furthermore, the differences in syntax and semantic rules between C++ and Java may explain the variations in our results.

Additionally, the performance of the Llama2-7b and the CodeLlama-7b models also reminds us that even code-specialized models like CodeLlama may not exhibit better capabilities than their base models in some tasks. In our experiment, the CodeLlama-7b model did not outperform the Llama2-7b model on both datasets, which was an unexpected result. CodeLlama is a family of code-specialized models of Llama2. Thus, they weresupposed to achieve better performance in code-related tasks like flakiness classification. However, Fig 3 show they had the same performance at the beginning of the fine-tuning. However, they exhibited different performances during the fine-tuning. The Llama2-7b model appeared to learn more information from the training dataset and achieved better accuracy and F1 score on the C++ dataset. The reasons for this can be complex and beyond the scope of our work; we leave this for future research.

Flakiness classification is a code-related task for LLMs. For LLMs, classification tasks are simpler than code generation since code generation requires models with higher understanding capabilities. Thus, we choose the models with 7-billion parameters. Models of this size have sufficient reasoning capabilities, relatively low hardware requirements, and perform well on less complex tasks. In general, larger LLMs possess stronger reasoning abilities and better performance in code understanding. For example, the Llama2-13b model achieves a pass@1 score of 24.5 on grouped academic benchmarks. Meanwhile, the Llama2-7b model only achieves a pass@1 score of 16.8 [37]. The pass@k metric represents the probability that at least one of the top k-generated code samples for a problem passes the unit tests [42]. If possible, one can use larger models to achieve better classification results in practice.

Moreover, based on our results, we propose to use different LLMs to classify the flakiness in different programming languages. The promising performance of Mistral-7b provides a potential solution for classifying the flaky tests for C++ projects. For Java projects, the Llama2-7b model can be deployed to achieve better performance. In addition, the precisions and recalls of each category are presented in Table 5. The Llama2-7b model achieved a perfect score of 1.0 in the Time category for both metrics. In practice, if we are more concerned about a certain type of flakiness, then we can use the model that has the best performance in classifying this type of flakiness. For example, the Llama2-7b model is good at classifying **Time** flakiness, so we can use it to classify **Time** flakiness.

LLMs have great potential in text generation and code understanding. Training an LLM from scratch requires a significant number of computational resources; however, finetuning a pre-trained LLM is less resource-intensive and can lead to better performance on specific tasks. For example, fine-tuning the Mistral-7b model on our dataset took less than six minutes with the given hardware. The fine-tuned model classified the C++ flaky tests perfectly. This model can reduce the time engineers spend detecting and determining root causes, thereby significantly reducing resource consumption, especially for large software systems. RAG can also be used to improve LLMs' performance in flakiness classification. Currently, there are many LLMs specialized for code, they have developed a basic understanding of the test code. In this case,

RAG can achieve better performance with less effort in processing the data [43]. In future studies, we can explore the combination of RAG and fine-tuning to improve the performance of LLMs in flakiness classification.

In our study, we collected the C++ dataset from several open-source projects and applied data augmentation. The dataset provides valuable material for further research on C++ flaky tests. However, we acknowledge that the dataset is insufficient to solve all practical issues. More datasets in this field are needed to help us understand and deal with various scenarios. Although our results suggest that LLMs can learn meaningful patterns with the synthetic data, due to our random train-test splitting strategy, there is a possibility that the original tests in the training set influenced the classification of corresponding mutated tests in the evaluation set. This could have inflated the model's performance, as LLMs like GPT-4 have been shown to recognize mutations. Future work should adopt a better data splitting strategy to eliminate this issue. For example, we candivide the original data and the corresponding generated data into the same subset to avoid this.

# 6. THREATS TO VALIDITY

In this section, we discuss potential threats to the internal and external aspects of our study, as well as the construct validity of our study.

## 6.1. Internal Validity

The internal threats to validity are concerned with the C++ dataset construction process of our experiment. In our study, we collected the C++ flaky tests from open-source projects on GitHub. The flaky tests were manually categorized based on developers' comments about the root cause of the flakiness, as explained in Section 3. However, in some instances, the comments did not specify the flakiness categories, requiring manual classification by us. Consequently, it is possible that the flaky tests were assigned to the wrong label, which could introduce bias into our experiment. To mitigate this risk, we reviewed the test code before we assigned the label to a flaky test. Then, we categorized the flakiness based on the developers' comments and our understanding. This can reduce the possibility of assigning wrong labels to our data and ensure its quality. In our study, we applied data augmentation on the collected C++ flaky tests using GPT-4. To make sure GPT-4 worked as we expected and generated accurate examples, we also checked all the augmented tests. This additional step helped to prevent issues related to poor data quality in our study.

## 6.2. External Validity

The generalization of our approach is one of the threats to external validity. In previous works, most research investigated the flakiness categorization of a single programming language, which is insufficient for the whole industry. The results cannot show the performance of the models in different programming languages. To evaluate the generalization of the LLMs, in our study, we fine-tuned three LLMs to classify the flakiness in C++ and Java projects. Our finding shows that the Mistral-7b model can be used to classify some of the flakiness in C++ and achieve a perfect result, but it cannot achieve the same performance on the Java dataset. Similarly, the results only show the abilities of the selected three models on the Java and C++ datasets. We recognize that the models used in this study may not generate the same results in the other programming languages. The other threat to external validity is the dataset used in our study. We collected 55

C++ flaky tests from GitHub and applied data augmentation to them. We acknowledge that the limited size of our dataset may contribute to overfitting. Also, this dataset may not accurately reflect the true distribution of flaky tests across all C++ projects. In addition, in practical applications, the coding habits and styles used in some fields will differ from the datasets used in our experiments. As a result, the models trained on our datasets might exhibit different performance on other datasets, potentially yielding higher or lower results depending on the distribution of the new dataset.

### 6.3. Construct Validity

One potential threat to construct validity lies in the metrics used to evaluate the performance of the models. In our study, the dataset was imbalanced across different categories.As shown in

Table 3, we had only 17 samples in the Too restrictive range category, compared to 90 samples in the Async wait category. This uneven distribution could lead to inaccurate results. To alleviate this threat, we used the weighted average of F1 score, precision and recall instead of their macro average. Weighted average considers the size of each category, and large categories have a greater weight when calculating the overall result. This approach ensures a more accurate and representative evaluation of model performance across all classes.

## 7. CONCLUSIONS

In our study, we deployed the CodeLlama-7b, Llama-7b and Mistral-7b models to classify the flakiness of flaky tests in C++ and Java datasets. We also created a dataset consisting of 362 C++ flaky tests, categorizing their flakiness. Each model was fine-tuned on the C++ dataset and the Java dataset. To answer our research questions, we evaluated them on the test datasets and observed excellent results.

Our empirical evaluation demonstrates that our approach performs excellently on the C++ dataset and achieves competitive results on the Java dataset. On the C++ dataset, the Mistral model achieved the highest performance among all the models. However, the Mistral-7b model only attained a score of 0.85 for the F1 score on the Java dataset. This indicates that the Mistral model is more suitable for classifying flakiness in the C++ dataset than the Java dataset. In contrast, the performance of all models on the Java dataset was similar, with the Llama2 model slightly outperforming the others. The Llama2 model achieved an F1 score of 0.89, whereas the Mistral and CodeLlama models scored 0.85 and 0.86.

FlakyCat deployed CodeBERT and Few Shot Learning to classify flakiness on the Java dataset, achieving an F1 score of 0.73. For our work, all three models achieved a higher F1 score on the Java dataset. The Llama2-7b model obtained an F1 score of 0.89, which is the highest. On the C++ dataset, our models, particularly the Mistral-7b, showed excellent performance, outperforming other models, with 1.0 for all the metrics. Our study shows the ability of LLMs to classify the flaky tests in two mainstream programming languages. The models achieve excellent results on both datasets, especially the Mistral-7b model. It obtained an F1 score of 1.0 on our C++ dataset. And the Llama2- 7b model also achieved satisfactory results on the Java dataset. Our results find that different LLMs exhibit different abilities in classifying the flakiness in Java and C++ projects. The results suggest that it is promising to employ these models in the CD process, therefore reducing the resource consumed and making the CD process more efficient.

## ACKNOWLEDGEMENTS

This research has been carried out at the Software Centre and Linköping University. The computations & data handling were enabled by the supercomputing resource Berzeliusprovided by the National Supercomputer Centre at Linköping University and the Knut and Alice Wallenberg Foundation. We thank the discussion with Jose Antonio Hernández López.

## AUTHORS

**Xin Sun** received Master's degree from KTH, Sweden, and he did bachelor's degree in Autonomous Systems. Currently, he is pursuing his PhD in Software Engineering from Linköping University. His research interests include LLMs and Software testing.

**Daniel Ståhl** is software subject matter expert at Ericsson AB and an associate professor of software engineering at Linköping University, Sweden. He has a background of 15 years of large-scale software development in the industry. He received his PhD degree from the University of Groningen, The Netherlands, in 2017 and his MSc degree from Linköping University, Sweden, in 2007.

**Kristian Sandahl** is a professor in Software Engineering at Linköping University in Sweden. His passion is to research and teach about methods and tools for developing, operating, and maintaining large-scale software systems with the right quality and cost. Most of his research is done in cooperation with industry and most of the teaching is done in connection to project work. He considers empirical methods and experiments essential for understanding the complexities of leveraging software technology effectively.